\documentclass[amssymb,amsmath,floatfix,twocolumn,letterpaper,nature]{revtex4}
\usepackage{graphicx}
\usepackage{lastpage}
\usepackage{amssymb}
\usepackage[utf8]{inputenc}
\usepackage{siunitx}
\newcommand{\ket}[1]{|#1 \rangle}

\begin{document}
\title{Requirements for fault-tolerant factoring on an atom-optics quantum computer}

\author{Simon J. Devitt$^{1}{}\footnote{electronic address: devitt@nii.ac.jp}$}
\author{Ashley M. Stephens$^{1}$}
\author{William J. Munro$^{2,1}$}
\author{Kae Nemoto$^{1}$}
\affiliation{$^{1}$National Institute for Informatics, 2-1-2 Hitotsubashi, Chiyoda-ku, Tokyo 101-8430, Japan}
\affiliation{$^{2}$NTT Basic Research Laboratories, NTT Corporation, 3-1 Morinosato-Wakamiya, Atsugi, Kanagawa 243-0198, Japan}

\date{\today}

\maketitle


\textbf{Quantum information processing and its associated technologies has reached an interesting and timely stage in their development where many different experiments have been performed establishing the basic building blocks. The challenge moving forward is to scale up to larger sized quantum machines capable of performing tasks not possible today. This raises a number of interesting questions like: How big will these machines need to be? how many resources will they consume? This needs to be urgently addressed. Here we  estimate the resources required to execute Shor's factoring algorithm on a distributed atom-optics quantum computer architecture.  We determine the runtime and requisite size of the quantum computer as a function of the problem size and physical error rate. Our results suggest that once experimental accuracy reaches levels below the fault-tolerant threshold, further optimisation of computational performance and resources is largely an issue of how the algorithm and circuits are implemented, rather than the physical quantum hardware.}

The prospect of an entirely new industry based on quantum mechanics has motivated technological development and 
led to a much better understanding of the principals governing our universe at the atomic scale. 
For quantum technology, experimental progress has been pronounced \cite{G06,HA08,PLZY08,PMO09,BW08,P12,L12}.  
Not only has a fledgling industry based on quantum key distribution already emerged \cite{magiQ,quin, IDQ}
but many experimental groups now routinely demonstrate the ability to create, manipulate and 
read-out multiple qubits in multiple physical systems with increasingly higher accuracy \cite{LJLNMO10}.  
The goal of developing a commercially viable, large-scale quantum computer is now coming into view.  
Theoretical progress is also an essential part, and fault-tolerant quantum error correction techniques, a necessity to deal with imperfect physical components, have been refined substantially \cite{K05,B06,RHG07}.  The adaptation of these techniques 
to the physical restrictions of quantum hardware has led to multiple architecture designs, indicating a clear pathway towards future quantum computers \cite{TED05, MTCCC05, HGFW06, FTYSPW07, SJ08, DFSG08, MLFY10, JMFMKLY10, YJG10}.  
   

While a large-scale quantum computer is still years away, it is now possible to make qualitative and quantitative predictions about the performance and required resources of such a computer.   Some of the previous predictions consider hardware architectures based on specific physical systems \cite{MLFY10, JMFMKLY10,TMCCC06,S07,CMG09}, which is an essential aspect in resource analysis.  However omit a full prescription for executing the algorithms in question. Others consider promising error-correction codes and circuits, such as post-selection \cite{K05} and topological error correction \cite{RHG07}, yet do so without reference to particular architectures or applications.  Above the hardware device level, there are a number of layers of implementation needed to finally run an algorithm.  By careful choice of all technological elements and 
the integration of all layers of implementation, a complete analysis is now possible, which we present in this manuscript.


A full account of the resources required for fault-tolerant quantum computation must consider a number of factors. Each physical component in our computer suffers from errors, therefore an appropriate error correcting code must be chosen to be compatible with the physical restrictions of the hardware.  Physical error rates must be suppressed below the fault-tolerant threshold of the chosen code.  Next, the code restricts the set of logically encoded gates that can be directly applied to encoded data.  Each gate in the high-level quantum algorithm is then decomposed 
into a universal set of fault-tolerant primitives.  To realise these universal primitives, ancillary states and protocols are typically required to 
enact teleported gates that could otherwise not be directly applied to the 
encoded data \cite{DN06,EK09,BK05}.  Each of these steps increases the total qubit/time overhead 
and must be carefully integrated together in a way that all steps are counted.


The precise details of how resources must be calculated depend on the properties of the architecture in question, the techniques utilised for fault-tolerant error correction, and the desired algorithm.  In this work we will be utilising a topological error correction code implemented on a large three dimensional cluster state of qubits \cite{RHG07}.  This error correction technique, despite the fact that it is the preferred protocol in large scale architectures, has only been briefly studied 
in regards to \emph{how} a large scale algorithm is implemented.  
Translating an abstract quantum algorithm into the specific operations needed in the cluster , i.e. the development of a \emph{classical} compiler,  has only just begun.  This step is anticipated to have a direct impact on the physical resources needed for computation.  
Typically, estimates consider the number of required gates in the high-level quantum algorithm and the basic amount of ancillary space needed for additional fault-tolerant protocols \cite{TMCCC06,MLFY10,JMFMKLY10}. However these estimates provide only a partial analysis.  Error correction codes inevitably suffer from constraints that need to be taken into account: 
specifically, the interaction of qubits required by the actual algorithm and qubits needed for ancillary fault-tolerant protocols.
The scheduling and routing of these ancillary protocols is often overlooked when estimating resources and are likely to dramatically 
affect resource estimates.


By contrast, compatibility of the topological model to hardware architecture has been demonstrated \cite{DFSG08, MLFY10, JMFMKLY10, YJG10}.  In our complete analysis, we will employ an atom-optics architecture \cite{DFSG08,DSMN11}, which is based on the photonic module \cite{DGOH07}.
The photonic module is a relatively simple device that allows an atomic qubit to mediate the generation of photonics entanglement.  The 3D cluster state to support topological error correction will then be created by an array of these devices.
Decomposition of each logical gate into a series of physical operations in this architecture is clear, and hence all the geometry and connectivity constraints at the logical and physical level can explicitly be included in the analysis.

The desired algorithm, Shor's algorithm, is a comparatively simple application compared other problems solvable by a quantum computer \cite{CMG09,J12}.  More importantly, it has a rich history of theoretical development and explicit circuit constructions.  Hence we can choose an  circuit construction amenable to the system design defined above.  However, to run the circuit, we still have to take the geometric constraints at the logical level into account.  Even though scheduling analysis at the physical level is taken care of by the topological quantum computer model, scheduling and arrangement of gates and ancillary operations within the logical space created by the topological cluster 
impact performance.  This step is largely unexplored, and leaves huge room for optimisation.  
We should remember that circuit optimisation needs to be done with such restrictions in mind.  The ability of an error corrected system to realise the optimal circuit size at the logical-level is dependent on adapting to these constraints, hence estimates should be made with care.


With this given computational system, the number of photonic modules and the time required to execute the algorithm as a function of the problem size and physical error rates desirably characterize  the computer.  As it is designed, the analysis explicitly deals with all aspects of the error corrected algorithm from the bottom device layer to the top abstract algorithm, giving a unique, but standardized estimation method.

\section{Preliminaries} In the topological cluster state model a three-dimensional cluster forms the effective Hilbert space in which computation takes place \cite{RHG06,RHG07}.  The photonic cluster state is continuously prepared from non-entangled photons by the hardware.   


Logical qubits are introduced as pairs of defects in the cluster.  Defects are created in the cluster by measuring physical qubits that 
define the defect in the $Z$ basis \cite{RHG07}. An entangling gate is realised by braiding pairs of defects. Logical errors occur when chains of physical errors connect or encircle defects, which is made less likely by increasing the circumference of defects and by increasing their separation. Physical qubits in the bulk of the cluster, those not associated with defects, are measured in the $X$ basis. This reveals the endpoints of chains of errors, from which the most likely set of errors can be inferred.  
To estimate physical resources, we are ultimately interested in the size of the three-dimensional cluster state required to execute Shor's algorithm. 

\begin{figure}[b!]
\begin{center}
\resizebox{60mm}{!}{\includegraphics{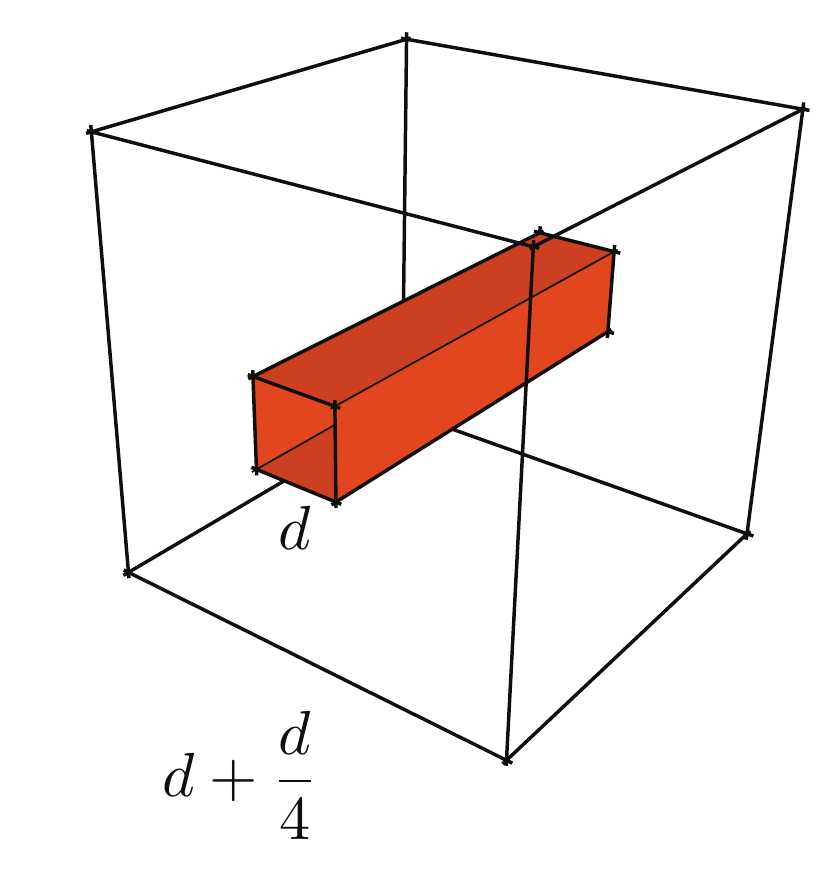}}
\end{center}
\vspace*{-15pt}
\caption{A logical cell; an error correction independent measure of the size of topological quantum circuits. 
The lengths are expressed in terms of unit cells of the cluster state. The 
qubit defect is the coloured region centred within the cell.}
\label{fig:cell}
\end{figure}

As the algorithm is executed at the logical level, it is useful to introduce a scale factor that essentially encapsulates the overhead associated with error correction \cite{RHG07}. A logical cell is defined as a three-dimensional volume of the cluster that has an edge length of $d+d/4$ unit cells, where $d$ is the distance of the error-correction code. Defects have circumference of $d$ unit cells and are separated by $d$ unit cells [Fig. \ref{fig:cell}].

\subsection{Shor's Algorithm} We now turn to the circuit for Shor's factoring algorithm. There are a number of different circuit implementations of the algorithm~\cite{VBE96,G98++,Z98,B03}, which assume that arbitrary sets of qubits can be simultaneously entangled without any penalty related to their separation.  In the topological model, as gates are realized by braiding defects, one could implement a gate over a long distance without any penalty.  
However, multiple gates are typically implement at the same time step and necessary scheduling within 
the topological cluster yields nontrivial overhead.  

An easier approach is to modify an existing circuit so that it only requires nearest-neighbour gates in some restricted geometry.  This can be done by adding \textsc{swap} gates to the circuit \cite{FDH04, MI05, K06}.  The Beauregard circuit \cite{B03,FDH04}, which we employ in this manuscript, is a Linear Nearest-Neighbour (LNN) construction \cite{FDH04}.  This circuit is not as efficient as others, but its explicit LNN construction means we can apply it directly to the topological cluster without further modification.  
With logical qubits arranged in a line, the circuit to factor an $L$-bit number requires $Q = 2L$ qubits and has depth $K = 32L^3$, to the leading order. The circuit is not inherently robust to errors~\cite{DFH06}, requiring an error rate per gate approximately, $10^{-1}/KQ = 10^{-1}/64L^4$, ensuring a 90\% chance of success.

%

\subsection{Gate decomposition} 
As with all error corrected models of quantum computation, not all gate operations can be directly applied in a fault-tolerant manner. 
At the logical level, only preparation of the states $|+\rangle$ and $|0\rangle$, $X$ and $Z$ gates, measurement in the $X$ and $Z$ bases and the \textsc{cnot} gate 
can be directly applied.  \textsc{swap} gates are achieved by deforming the trajectory of the defects with which they are associated. To complete a universal set we add the $R_z(\pi/8)$ and $R_z(\pi/4)$ rotations \cite{RHG07}. To apply these gates we perform a teleported gate using the ancillary states $|A\rangle = (|0\rangle + e^{i\pi/4}|1\rangle)/\sqrt{2}$ and 
$\ket{Y} =( |0\rangle + i|1\rangle)/\sqrt{2}$.  Each time we attempt the $R_z(\pi/8)$ gate, there is a 50\% chance that 
a $R_z(\pi/4)$ correction is required.   

To ensure that the error rate of the $R_z$ rotations are sufficiently low, the states $|A\rangle$ and $|Y\rangle$ must be of sufficient fidelity. 
As these ancillary states are prepared in the cluster via injection protocols \cite{RHG07}, state distillation is used to increase the fidelity of the ancilla state \cite{BK05}, consuming multiple $|A\rangle$ or $|Y\rangle$ states with a lower fidelity.   This process can 
be concatenated until the desired fidelity is reached. If $p_l$ is the error probability on the state after $l$ levels of state distillation, then $p_{l+1}^A = 35(p_l^A)^3$ and $p_{l+1}^Y = 7(p_l^Y)^3$ for $|A\rangle$ and $|Y\rangle$ respectively \cite{BK05}. Each distillation circuit is probabilistic with a failure probability of $O(p)$.  

Given our set of logical gates, which now includes the $R_z(\pi/8)$ rotation, we need to decompose the circuit for Shor's algorithm into a sequence of these fault-tolerantly implemented gates. For an upper bound on the number of gates needed we will (pessimistically) assume 
every gate is a non-trivial phase rotation that must be approximated by a sequence of logical gates found using the 
Solovay-Kitaev algorithm~\cite{DN06}, and each gate in this sequence is the $R_z(\pi/8)$ rotation, which is most resource intensive amongst our logical gates constitutng over 50\% of the decomposition \cite{F04}.  Numerical results suggest that a sequence of $\Lambda = 19.6\log(1/\epsilon)-10.5$ gates is required to achieve an arbitrary single qubit rotation with accuracy $\epsilon$~\cite{F04}. Hence, to achieve the required error rate, each logical gate in the circuit can be estimated as a sequence of $\Lambda = 19.6\log(640L^4)-10.5$ gates.  

\section{Results} 

\subsection{Braided circuits} 
\begin{figure*}
\begin{center}
\resizebox{160mm}{!}{\includegraphics{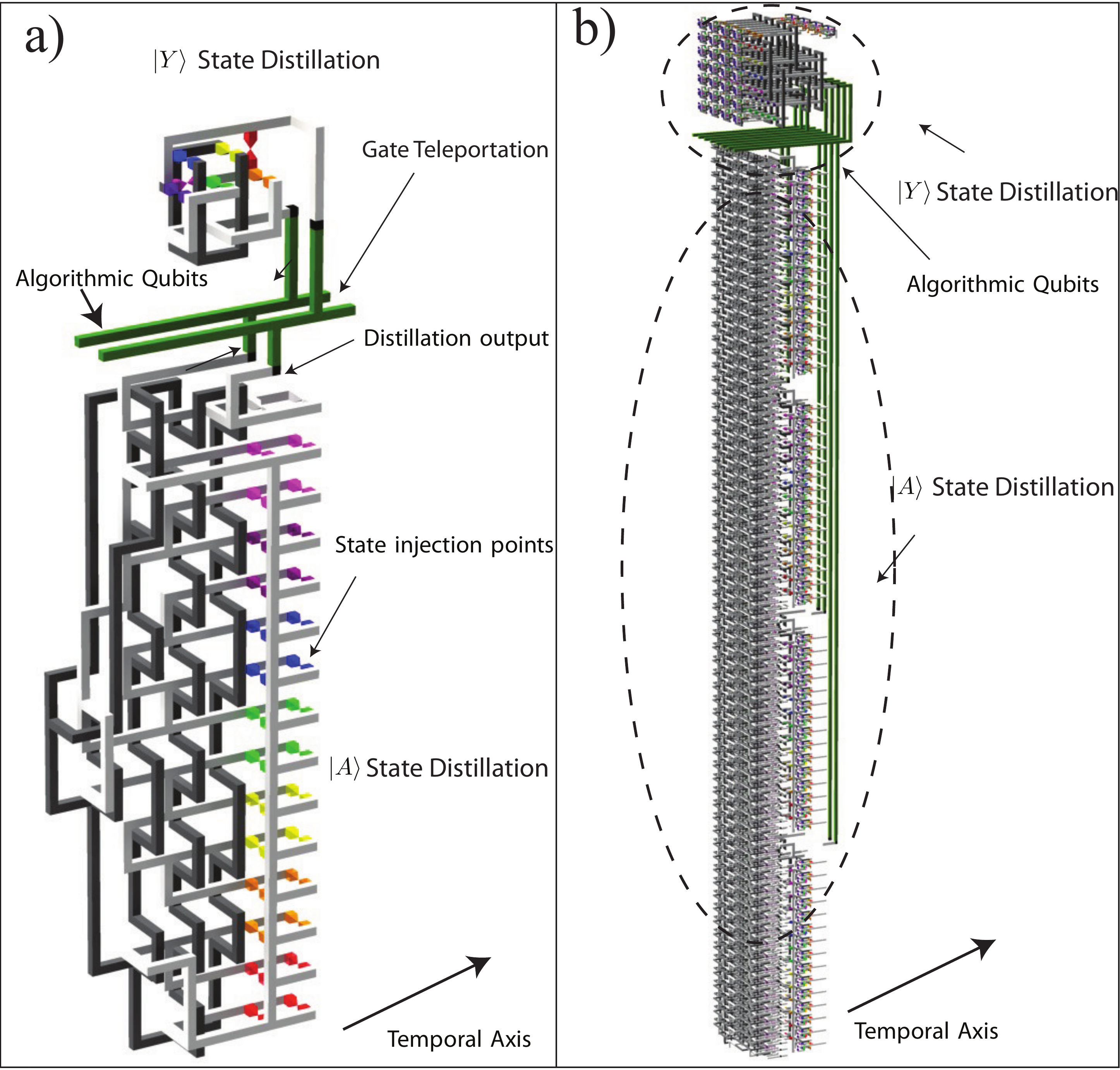}}
\end{center}
\vspace*{-15pt}
\caption{Explicit braiding constructions for a $R_z(\pi/8)$ rotation in the topological cluster at \textbf{a) one} and 
\textbf{b) two} levels of concatenated state distillation.  The temporal axis in the cluster is illustrated.  For a 
detailed explanation of these constructions see the supplementary material.  Qubits that are part of 
the algorithmic circuit for Shor are illustrated in green.  The cluster volumes and depths 
for these two circuits are $V = \{210, 1386\}$ and $D = \{5,9\}$ respectfully.  Each sequence is designed such 
they can be stacked together efficiently in either the temporal or spatial directions in the cluster.}
\label{fig:Tgate}
\end{figure*}

We now translate the decomposed circuit for Shor's algorithm to a sequence of braids in the three-dimensional cluster state. As each gate in the 
algorithm is assumed to be a $R_z(\pi/8)$ rotation, this is the logical gate that will be designed.  Shown in Fig. \ref{fig:Tgate} is 
the braiding sequence for the logical $R_z(\pi/8)$ rotation at one and two levels of concatenated state distillation.  
Full details of these gate constructions are detailed in supplementary material.  The braiding sequence is 
compressed manually into a cuboid such that they can be stacked tightly in the spatial and temporal directions in the cluster. 
The algorithmic qubits (the ones specified in the Beauregard 
circuit) are the green defects (two defects per algorithmic qubit occupying a cross sectional area of two logical cells).  Immediately above each 
algorithmic qubit is an empty region of the cluster, this empty space is utilised for braided logic and \textsc{swap} gates 
required by the Beauregard circuit.  The linear nearest neighbour design of the original circuit 
ensures that no further optimisation is required at the algorithmic level and that the defect layout of algorithmic qubits in the cluster 
is sufficient to realise the depth of the original circuit.  Above this empty region is the distillation space for 
$\ket{Y}$ states, required to implement a $R_z(\pi/4)$ correction gate for each applied $R_z(\pi/8)$ gate and 
Hadamard operations.  Below the algorithmic qubits is the distillation space for $\ket{A}$ states.  

At one level of concatenation, each algorithmic qubit has a dedicated $\ket{A}$ and $\ket{Y}$ state distillery.  As the algorithmic 
layer is linear, these distilleries connect from above and below in the cluster (direct connections in the topological model 
correspond to teleported gates \cite{RHG07}).  

For two levels of concatenation the repeating cuboid encapsulates four algorithmic qubits.  
The first concatenation level has physical injection points for low fidelity $\ket{A}$ and $\ket{Y}$ states and the size of 
the defects are half of what is required at the algorithmic layer.  this reduced size and separation of defects for the first 
concatenation level is because distillation circuits have a residual error.  Therefore if the error of an injected state at the 
physical level is $O(10^{-3}-10^{-4})$, then implementing full strength error correction for these circuits is redundant.  The 
residual error from distillation will always dominate.  At the second layer 
of concatenation, the residual error becomes commensurate with the required logical error needed for computation. Therefore, 
after the first layer of concatenation, defects are expanded and separated to the same size as the required error correction for the 
algorithm.  Additionally, at the second level of concatenation, the state injection for corrective $\ket{Y}$ states, needed for the 
$\ket{A}$ state distillation, becomes level one $\ket{Y}$ state circuits, placed in the relevant free space in the cluster.  

The application of corrective $R_z(\pi/4)$ rotations for $\ket{A}$ state distillation and the probabilistic nature of the 
circuits themselves are compensated at the second level of concatenation by utilising free space to add extra distilleries [See 
supplementary material].  At the 
first level of concatenation, for $\ket{Y}$ states, there is sufficient space for one extra circuit adjacent to the 
second level circuit, to compensate for any one failure at level one.  For $\ket{A}$ state distillation there is space for 
two extra circuits within the cuboid to compensate for a given circuit failure.  These circuit failures 
occur with probability $O(p)$, with $p$ the fidelity of the injected states.  Given the extra space for spare 
level one circuits and assuming $p$ is $O(10^{-3} -10^{-4})$, we will 
have too many failures at level one with a probability $O(10^{-5} - 10^{-7})$ for $\ket{Y}$ states and 
$O(10^{-7} - 10^{-9})$ for $\ket{A}$ states.   Therefore, we expect that we will not have sufficient 
first level states every approximately $10^5$-$10^7$ Logical gates.  While these failures result in an increase 
in circuit depth, they occur infrequently enough to be neglected.  Finally, a total of 15 first level circuits for 
corrective $\ket{Y}$ states, needed by the second level $\ket{A}$ state circuit, are used.  The probability that 
not enough level one $\ket{Y}$ states are available is given by $\frac{15p}{2^{15}}$ (i.e. all 15 
$R_z(\pi/8)$ corrections are needed and a single level one $\ket{Y}$ distillation failure occurs), and if 
$p$ is $O(10^{-3})$ this is also of $O(10^{-7})$.
Corrective $\ket{Y}$ states, needed with a probability of 0.5 for the logical $R_z(\pi/8)$ gate are located above the 
algorithmic layer.

The total logical volume of cluster for one and two levels of state distillation can be calculated explicitly.  For one 
level of concatenation, each $R_z(\pi/8)$ gate occupies a volume of $V = 5\times 21 \times 2$ cells with a depth along 
the temporal axis of the cluster of $D=5$ cells and a cross sectional area of $A=21\times 2$.  
For two levels of concatenation the volume is $V = \frac{8\times 77 \times 9}{4} = 1386$ 
where the factor of four accounting for the fact that the cuboid represents four gates.  The number of cells along the temporal 
axis is $D=8$ and a cross sectional area of $A = 77\times 2$.

\begin{figure}
\begin{center}
\resizebox{83mm}{!}{\includegraphics{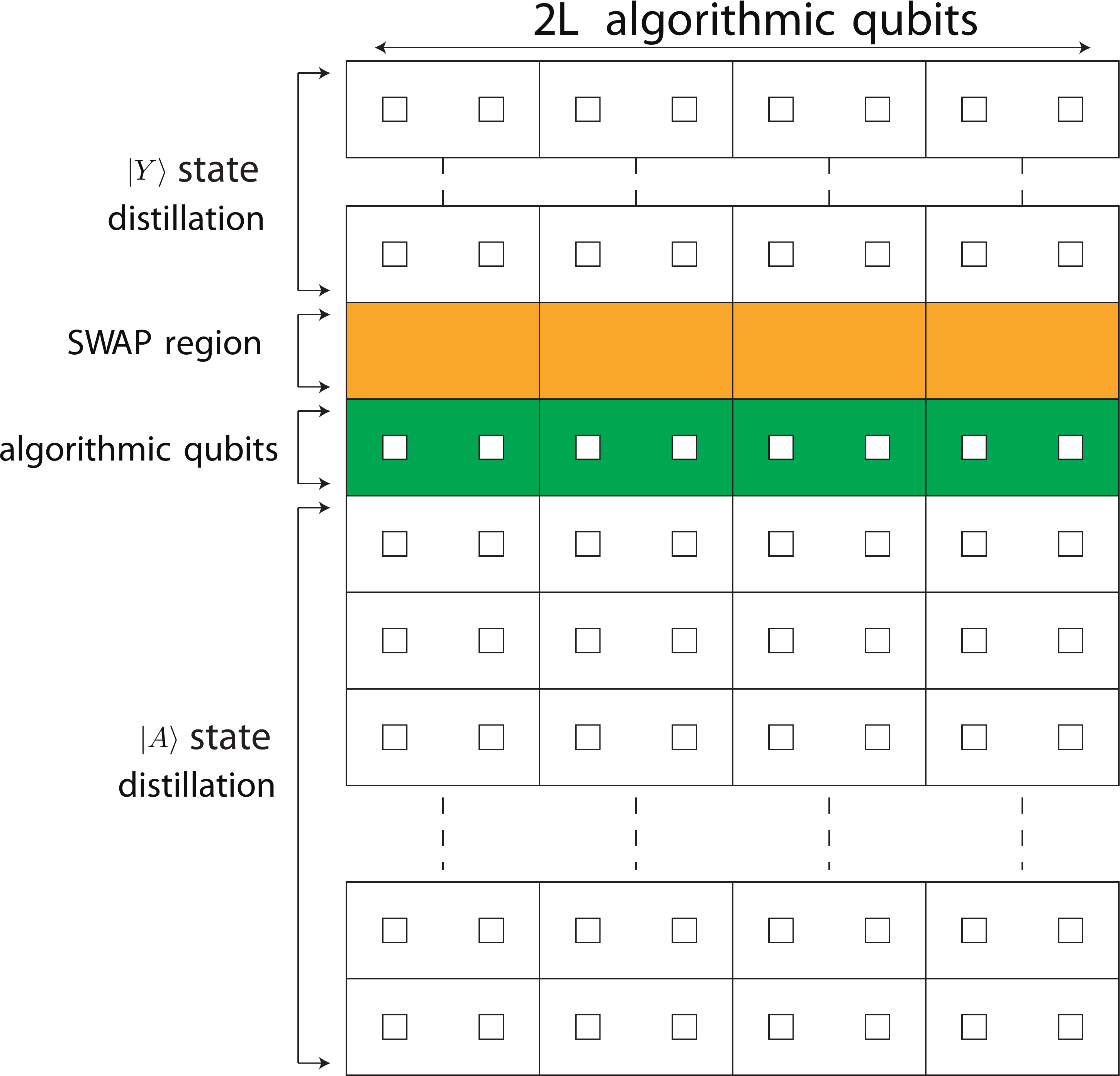}}
\end{center}
\vspace*{-15pt}
\caption{Layout of logical qubits for Shor's algorithm,including $\ket{A}$ and $\ket{Y}$ distillation.}
\label{fig:layout}
\end{figure}

\subsection{Cluster Volume} To determine the total size of the cluster state, we need to know the amount of error correction and state distillation required. Each logical gate requires $\Lambda \times V$ logical cells.  Hence, the failure probability of such a gate needs to be,
\begin{equation}
\begin{aligned}
1-(1-p_f)^{\Lambda V} \leq \frac{1}{640L^4},
\end{aligned}
\end{equation}
where $p_f$ is the error rate of a logical cell and the right hand side sets the target error rate for gates in the circuit for Shor's algorithm. 
For standard depolarizing noise, we 
can estimate the failure of a single \textit{logical} volume of the cluster as, $p_f \approx C_1\left(C_2 p/p_{th}\right)^{\lfloor(d+1)/2\rfloor}$, where $d$ is the distance of the code, $p$ is the physical error rate, $p_{th}$ is the threshold error rate, which is estimated to be approximately 0.62\%  and $C_1 \approx 0.13$ and $C_2 \approx 0.61$ \cite{RHG07,BS10}.  
Assuming that $p_f\ll1$ and $1/640L^4\ll 1$, the distance required to achieve the target error rate is
\begin{equation}
d \geq \bigg{\lceil} \frac{2\log\left(640C_1L^4\Lambda V \right)}{\log(p_{th})-\log(C_2 p)} - 1\bigg{\rceil}.
\label{eq:dist2} 
\end{equation}
Here we assume that the residual error after state distillation is below the error rate of a logical cell, such that $7^{(3^l-1)/2}p^{3^{l}} \leq p_f$ and $35^{(3^l-1)/2}p^{3^l} \leq p_f$ for $|Y\rangle$ states $|A\rangle$ states respectively. 
These conditions determine the level of state distillation required.  
Only for very large $L$ or for high values of $p$ does state distillation require a maximum of three concatenated levels.  
The volume and depth at this level 
was extrapolated from the level two circuits at $V=10000$ and $D=15$.  

Finally, we can specify the properties of the entire cluster state. 
The cluster contains $4L \times A$ logical cells.  The total cross-sectional area of the cluster is $5Ld\times 5dA/4$ physical unit cells. The third dimension of the cluster represents 
the temporal axis and its size determines the computational time. The depth of a single logical gate is $\Lambda\times D$ and the depth of a single $R_z(\pi/8)$ gate is $\Lambda D\times (5d/4)$. Therefore, the total depth of the cluster is $(32L^3\Lambda D)\times(5d/4)$.

\subsection{Physical Resources} 
In the architecture, photonic modules are used to prepare the cluster state and also to initialize and measure single photons \cite{DSMN11}.  
There is a one-to-one mapping between the cross sectional size of the 3D cluster and the number of required modules. For a cluster with a cross-sectional area of $N_1\times N_2$ physical unit cells, a total of $(2N_1+1)(2N_2+1)$ optical lines are present, half require two modules for photon detection and 
half require four.  All optical lines require one module as a probabilistic source. The number of modules required to prepare the cluster state is $2(N_1+2)(N_2+1)+2(N_2+2)(N_1+1)$~\cite{DFSG08}. This gives a total number of modules equal to 
$\left(12+14N_1+14N_2+20N_1N_2\right)$, 
with $N_1 = 5Ld$ and $N_2 = 5d/4A$. In addition to the number of modules, we can specify the physical size of the computer and its runtime. The dimensions of the computer are $S_x = 5LdM$ and $S_y = 5dMA/4$, where $M \times M$ is the surface area of a photonic module (with depth $< M$) \cite{DFSG08}. The physical depth of the computer is $S_z \leq 2Tc_f$, where $c_f$ the speed of light in fiber. This depth is governed by the optical lines that recycle photons from the detectors to the sources \cite{DSMN11}. The time required to run the algorithm is 
$32L^3\Lambda D\times5d/4\times 2T$, where $T$ is the time required to prepare a single layer of the cluster state~\cite{DFSG08}, 
corresponding to the operational speed of the photonics module. 

\begin{figure*}
\begin{center}
\resizebox{180mm}{!}{\includegraphics{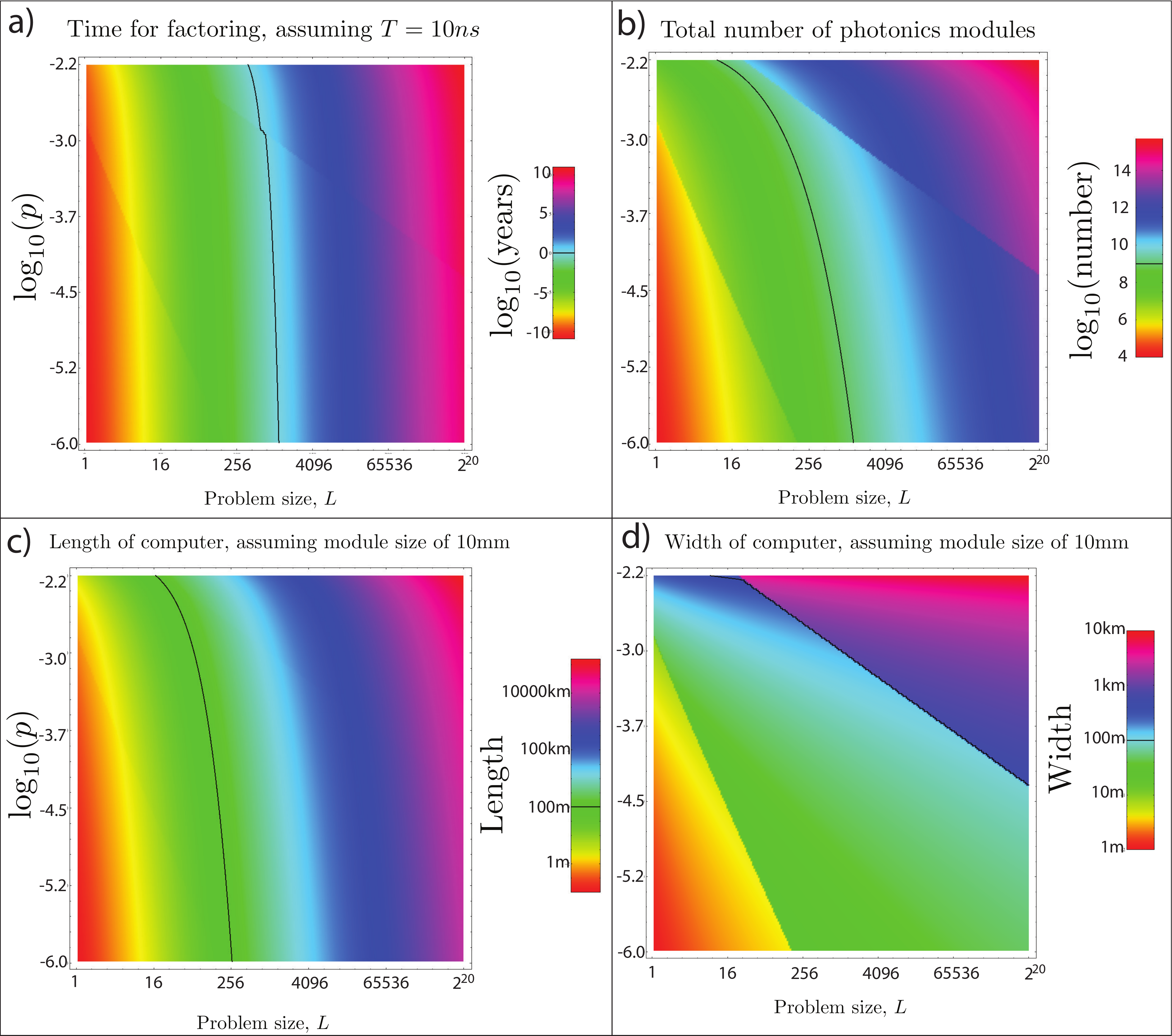}}
\end{center}
\vspace*{-13pt}
\caption{({\bf a)}) Time, ({\bf b)}) number of photonic modules and ({\bf c), d)}) computer size required to factor an $L$-bit number with an error rate $p$.  The discontinuities represent points where the concatenation level for state distillation increases.}
\label{fig:results}
\end{figure*}

Figure \ref{fig:results} shows the runtime of the algorithm, the total number of photonic modules and the dimensions of the computer as functions of the physical error rate and the problem size. Here we have assumed that $p_{th}=0.62\%$ \cite{RHG07,BS10}, $M = 10 $mm and $T = 10$ ns \cite{SGMNH08}.  Contour lines in Figure.~\ref{fig:results} indicate where the time to completion is one year, when the total number of photonic modules is 1 billion, and when the cross sectional dimensions are 100m.  With an error rate an order of magnitude below $p_{th}$, the largest problem size that can be completed within a year is $L\approx 820$.

\section{Discussion} 
The current record for factoring general integers is $L=768$ \cite{K10}, hence as anticipated, these results 
show the superiority of quantum computation.  However, at the same time, they seem not to demonstrate a significant 
increase in the processing power of quantum computers.   

Our results give a comfortable upper 
bound for the resource requirements using explicit constructions in the topological model.  The time required to factor 
a 1024-bit number in this analysis is 2.15 years with 1.9 billion photonic modules, required to prepare the 
cluster.  An interesting question to ask here might be how these numbers can be compared with the fundamental circuit used in this analysis.  The basic circuit 
requires a computational depth of $32L^3$ and $2L$ qubits.  For physical gate times of 
10ns, for $L=1024$, the error 
correction overhead is $2.3\times10^{7}$ temporally and $9.4\times10^{5}$ in terms of qubits/modules.  
These numbers are based on a physical error rate an order of magnitude below 
threshold.   This overhead, resulting from the error correction, can potentially be significantly reduced by optimisations unrelated to the fundamental hardware.  This can be easily highlighted by the fact that decreasing the error rate by an order of magnitude results in a speed-up of to 1.14 years.  The same speed-up can be achieved by compactifying the topological circuits shown here by 44\% along the temporal 
axis of the cluster.


There has been many other resource estimates made for a computer employing both concatenated and topological coding models.  Thaker \textit{et al.}~estimated that to factor a $1024$-bit number using an architecture based on trapped ions would take around 25 days~\cite{TMCCC06}. Van Meter \textit{et al.}~estimated a $2048$-bit number on a distributed architecture based on quantum dots would take around 400 days~\cite{MLFY10}. Jones \textit{et al.}~recently improved the latter estimate to around 10 days \cite{JMFMKLY10} by utilising a monolithic array of dots and increasing the speed of fundamental error correction cycles.  New results in superconducting designs suggest a factoring time, for a 2000-bit number, slightly less than one day \cite{FM12}.  In these estimates differences arise due to how the algorithm is implemented.   
Until a complete analysis is performed, it is meaningless to directly compare them.  In particular, more resource efficient techniques are utilised in these results which need to be explicitly integrated within the topological model for future estimates.



All resource estimates, including ours, illustrates that large fraction of the overhead arises from the need to prepare ancillary states.  Other results assume sufficient space within the computer such that ancillary protocols 
can be completed rapidly enough that the depth of the algorithmic circuit is unchanged.  This could be of significant 
benefit.  However, the appropriate routing of these ancillary protocols need to be explicit.   How distillation 
circuits are interfaced with data qubits needs to be detailed and exactly which protocols are utilised needs 
to be analysed.  Estimates from Refs. \cite{JMFMKLY10,FM12} use the most optimal circuit for Shor's 
algorithm \cite{VBE96,CDKM04}.  
This circuit has not yet been adapted to the geometric constraints of the topological cluster.  Until an appropriate 
construction is presented for the topological cluster it is difficult to assume that the circuit size will remain unchanged.  If such a circuit design is presented, then we anticipate immediate reductions in resources.   
Previous results also assume that various subcomponents of a fault-tolerant implementation can be applied 
without space/time penalty.  There has been many results published optimising various components in a 
fully error corrected quantum algorithm \cite{FD12,F12+,J12+,BH12}.  However, each of these results have been derived in isolation, 
some have not been converted into the topological model and none have been carefully integrated together.  This is the primary challenge 
of topological computation.  Subcomponents may be efficient, but the success of a large-scale computation 
requires delicate integration.  Our results illustrate that there is a significant gap between optimistic resource 
estimates and those performed using explicit circuit contractions.  


It is clear that before a quantum computer is actually build that algorithmic compilation is a necessity.  Reducing the burden on experimental development is ultimately a function of how we realise abstract algorithms.   This analysis illustrated that there is much work to be done.  While the topological model is promising, its ultimate success is dependant on continual efforts to integrate all necessary protocols in a way that minimises the number 
of devices and the time required to execute an algorithm.  

\section{Acknowledgements} We thank N.~C.~Jones and A.~G.~Fowler for helpful discussions. This work was partly supported by the 
Quantum Cybernetics (MEXT) and the FIRST project in Japan.


\bibliography{bib1}
\bibliographystyle{nature}

\section{Appendix}
Here we detail the component constructions required to build a logical $R_z(\pi/8)$ gate, which forms the basis for the 
resource estimates illustrated in the main text.  The goal of these braid constructions is to minimise the total logical 
volume to implement the fault-tolerant gate and to ensure that braids are constructed in a manner compatible with the underlying 
circuits and in such a way that they can be packed densely within the overall topological cluster.  
The techniques used in this section to achieve compact braiding utilises results from Ref. \cite{RHG07} and Ref. \cite{FD12}.

\section*{A. Primitive operations}

\begin{figure*}[ht!]
\begin{center}
\resizebox{150mm}{!}{\includegraphics{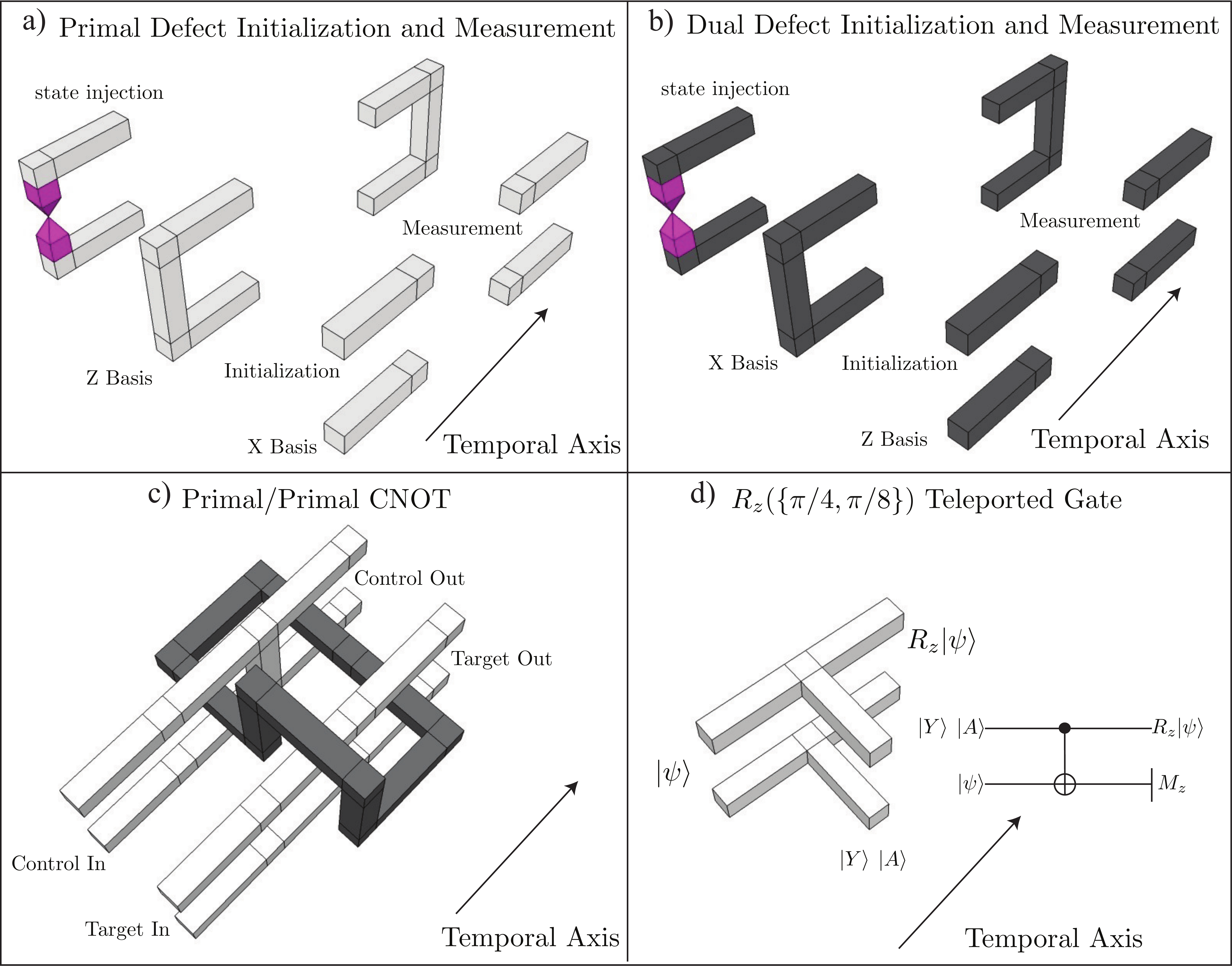}}
\end{center}
\vspace*{-15pt}
\caption{Examples of basic operations that are used to construct braid sequences. {\bf a)} Primal defects in a horseshoe shape are used to prepare a logical qubit in the state $|0\rangle$ and to measure a logical qubit in the $Z$ basis. An arbitrary state can be prepared by measuring one of the physical qubits (shown in pink) in a rotated basis as the defects are created. {\bf b)} Similarly, dual defects in a horseshoe shape are used to prepare a logical qubit in the state $|+\rangle$ and to measure a logical qubit in the $X$ basis. {\bf c)} A \textsc{cnot} gate can be achieved by braiding a pair of dual defects (prepared in the state $\ket{+}$) with three pairs of primal defects (the control qubit, the target qubit and an extra qubit prepared in the state $\ket{0}$)  \cite{RHG07}.  {\bf d)} A teleported $Z$ rotation can be achieved by attaching the relevant ancillary state to the data qubit \cite{RHG07}.}
\label{fig:legend}
\end{figure*}

First let us introduce the primitive fault-tolerant operations that are allowed in the topological model.  We introduce 
five types of gates; measurement, initialisation, state injection, 
the two-qubit \textsc{cnot} and the teleported phase rotation, $R_z(\theta)$, $\theta = \{\frac{\pi}{4}, 
\frac{\pi}{8}\}$.  These are illustrated in Fig. \ref{fig:legend}. 

\section{Compactified distillation circuits}
\begin{figure}[ht!]
\begin{center}
\resizebox{60mm}{!}{\includegraphics{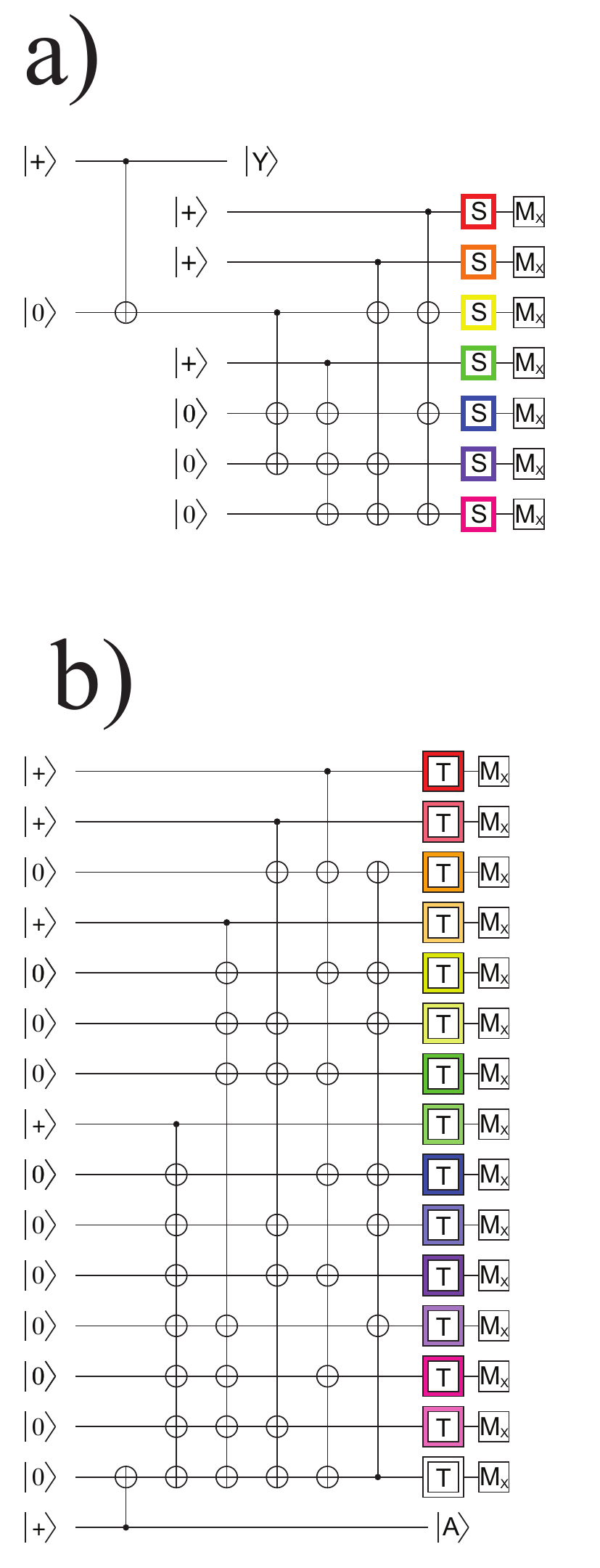}}
\end{center}
\vspace*{-15pt}
\caption{Quantum circuits required for distillation of the states \textbf{a)} $\ket{Y}$ and \textbf{b)} $\ket{A}$.  
The coloured boxes represent where error prone states are injected into the cluster.  This solar coding corresponds 
to the coloured injection points in braiding diagrams throughout this paper.}
\label{fig:circuits}
\end{figure}

\begin{figure*}[ht!]
\begin{center}
\resizebox{150mm}{!}{\includegraphics{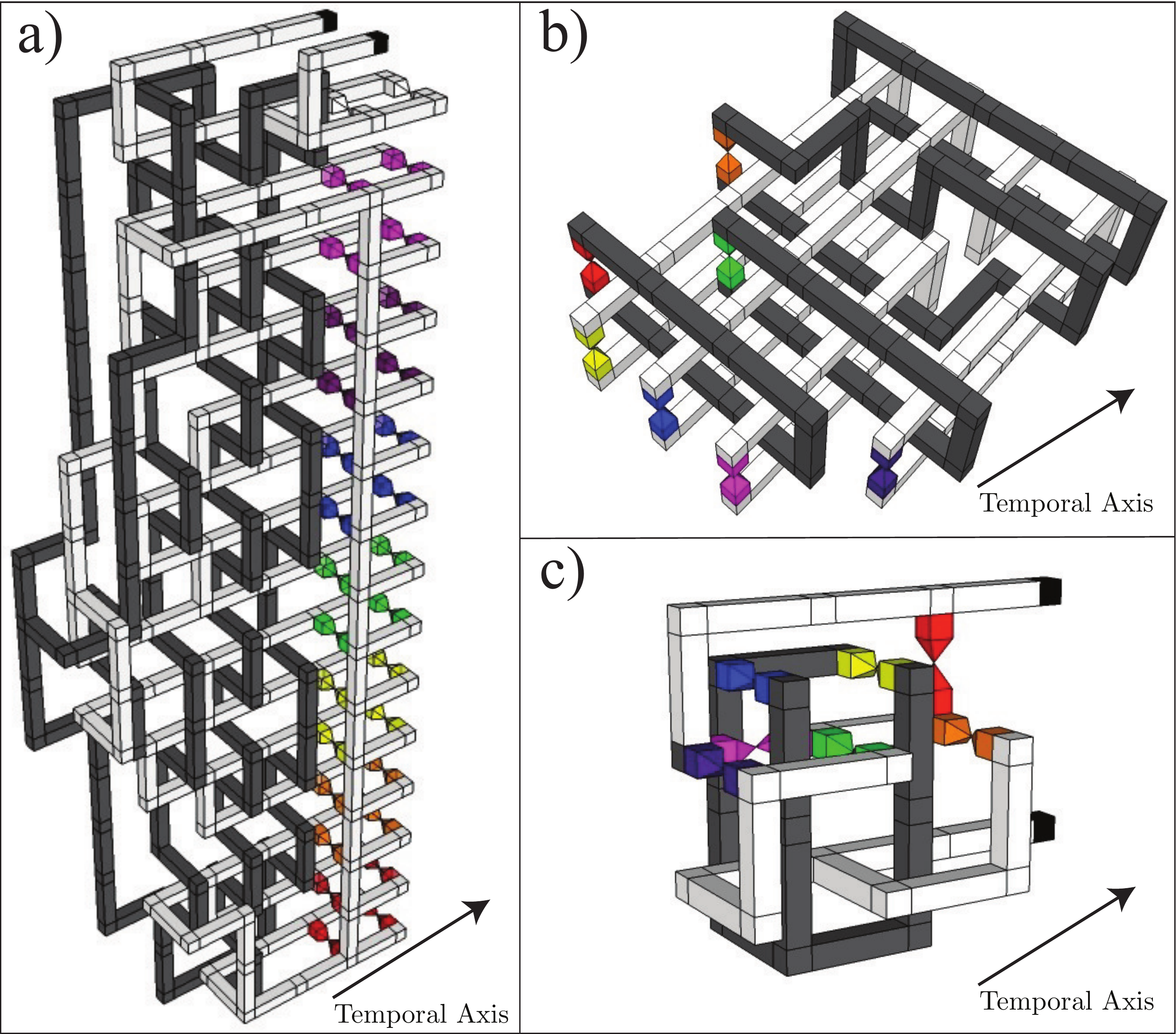}}
\end{center}
\vspace*{-15pt}
\caption{Braiding diagrams for \textbf{a)} $\ket{A}$ state distillation and \textbf{b), c)} $\ket{Y}$ state 
distillation.  These compactified circuits are from Ref. \cite{FD12}.  We illustrate two designs for 
$\ket{Y}$ state distillation as the circuit in Fig. B) will be utilised at the second concatenation level. Coloured 
pyramid structures represent state injection points to implement the $R_z(\pi/4)$ and $R_z(\pi/8)$ gates 
needed in the distillation circuits.}
\label{fig:braid1}
\end{figure*}

With these primitive operations, the standard circuits for $\ket{A}$ and $\ket{Y}$ state distillation can be converted to a compactified braiding 
sequence.  Shown in Fig. \ref{fig:circuits} are the canonical circuits for distillation.  For the $\ket{Y}$ state, one half of a logical Bell pair is 
encoded with the $[[7,1,3]]$ error correcting code, after which a transversal $R_z(\pi/4)$ gate is applied to the 
encoded half and measured in the $X$ basis.  For $\ket{A}$ state distillation half of the Bell pair is encoded with the $[[15,1,3]]$ Reed-Muller 
code and a transversal $R_z(\pi/8)$ gate applied prior to logical measurement.  Given the correct set of measurement results a purified 
copy of the $\ket{Y}$ or $\ket{A}$ state is teleported to the output half of the original Bell state. 
In Ref. \cite{FD12} it was shown how the canonical versions of the braided logic can be compactified using previously known techniques and 
a process known as defect bridging.  Illustrated in Fig. \ref{fig:braid1} are the compactified versions of the circuits shown in Fig. \ref{fig:circuits}.
In Fig. \ref{fig:braid1}a) we show the compactified version of the $\ket{A}$ state distillation circuit.  The sets of coloured pyramids represent the 
injection and gate teleportation needed to realise the transversal $R_z(\pi/8)$ and corrective $R_z(\pi/4)$ gates applied to the 
encoded half of the initial Bell state, with colour coding matching Fig. \ref{fig:circuits}.  
Imbedded within the defect structure is the logical $Z$  measurement present in the 
teleported gate circuit and this logical measurement result dictates if a further corrective $R_z(\pi/4)$ rotation needs to be applied (again 
via injection and teleportation, this time using a $\ket{Y}$ state).  For this circuit, a strictly enforced temporal axis is needed because the 
logical measurement of the first injection and teleportation dictates if the second one needs to be applied.  

In Fig. \ref{fig:braid1}b) and c) we show the compactified version of the $\ket{Y}$ state distillation circuit.  Fig. \ref{fig:braid1}b) illustrates the compact version, 
using known techniques \emph{excluding} bridging, while Fig. \ref{fig:braid1}c) illustrates the final version after defect bridging.  
We present both circuits as they 
will both be used in a concatenated distillation sequence.  Unlike the $\ket{A}$ state distillation circuits, there is no strict enforcement of a 
temporal axis in the cluster.  Although the transversal $R_z(\pi/4)$ operation for $\ket{Y}$ state distillation is also probabilistic for each 
teleported gate, the correction operation is a simple $Z$ gate which can be applied via appropriate classical tracking of the Pauli frame.  
In all three versions of the circuit, the relevant output defects are shown with black caps. 

\section{Level 1 concatonated gate}

The level one concatenated $R_z(\pi/8)$ gate is relatively simple to construct, primarily because all injection points correspond to 
physical qubits in the topological cluster.  For each algorithmic qubit in the computer, the region below is devoted to $\ket{A}$ state 
distillation while one logical cell above contains empty cluster to enable \textsc{swap} and \textsc{cnot} operations between 
algorithmic qubits.  The region above this layer is devoted to $\ket{Y}$ state distillation.  Unlike higher levels of 
concatenation, a single $R_z(\pi/8)$ gate can be defined which can be repeated along the spatial and temporal axes of the cluster. 
\begin{figure*}[ht!]
\begin{center}
\resizebox{150mm}{!}{\includegraphics{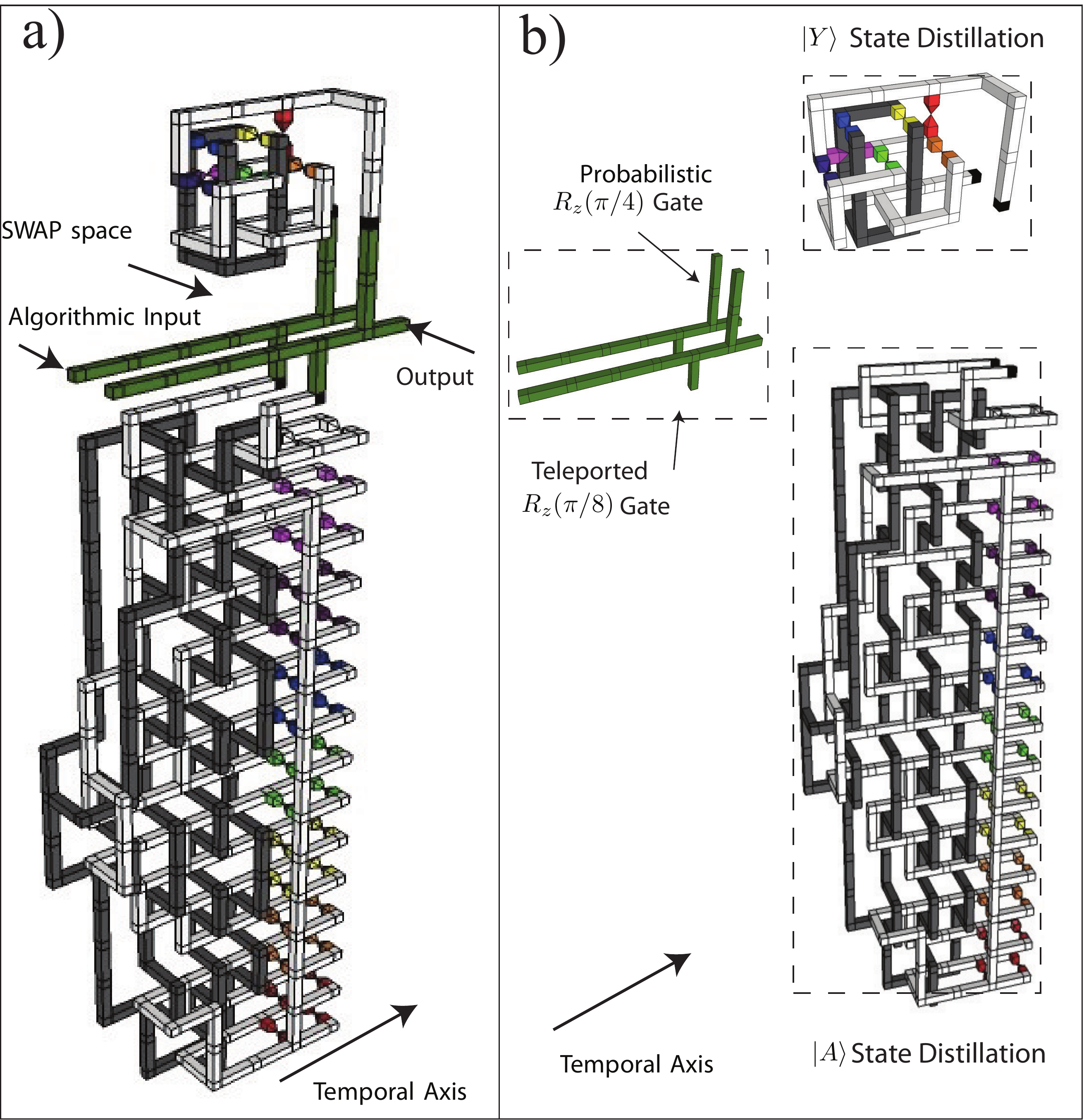}}
\end{center}
\vspace*{-15pt}
\caption{$R_z(\pi/8)$ gate at one level of concatenated distillation. Fig. \textbf{A)} shows the connected 
circuit, including $\ket{A}$ state distillation, $\ket{Y}$ state distillation for the corrective $R_z(\pi/4)$ operation 
and the (green) algorithmic qubit.  Fig. \textbf{B)} illustrates each of the three components.  The temporal axis 
through the cluster is illustrated. }
\label{fig:1level}
\end{figure*}
Fig. \ref{fig:1level}a) illustrates the complete gate which has a depth along the temporal axis of $D=5$ and a cross sectional area in the 
cluster of $A = 21\times 2$.  The algorithmic qubit is idle until the distillation operations are complete and with a 50\% probability, the 
corrective $R_z(\pi/4)$ gate need not be applied.  At one level of 
concatenation, all defects in the circuit have the same size and separation.  
For a level one concatenated circuit there is no extra space for distillation circuits to 
compensate for a failure event in the circuit itself.  As the \emph{logical} error rate required by the computer at one level of concatenation is 
high (for an experimentally feasible physical gate error), the total number of gates per logical time step will be quite comparatively small 
and hence the probability of a failed distillation circuit \emph{per logical time step} is quite low.  
Therefore, in the even that a distillation circuit fails, this structure would be repeated.  In our analysis we assume that all gates are 
$R_z(\pi/8)$ rotations, however in reality this is not the case.  In the event of an occasional failure, a repeated distillation circuit can be performed 
in cluster spaces otherwise vacant due to other logic operations during computation. 

\section{Level 2 concatonated gate}
Forming a second level concatenated $R_z(\pi/8)$ gate is significantly more complex.  This is due to injection points at the second level 
coming from outputs of level one circuits.  From Fig. \ref{fig:circuits}a) you can see that accessibility to the $\ket{A}$ state injection points within 
the braiding structure is quite limited.  As a well defined temporal axis has to be maintained, 
we need to modify the $\ket{A}$ state circuit such that these injection points can be connected easily to the level one outputs.  The following 
sequence of images illustrates the deformations.  
\begin{figure}[ht!]
\begin{center}
\resizebox{40mm}{!}{\includegraphics{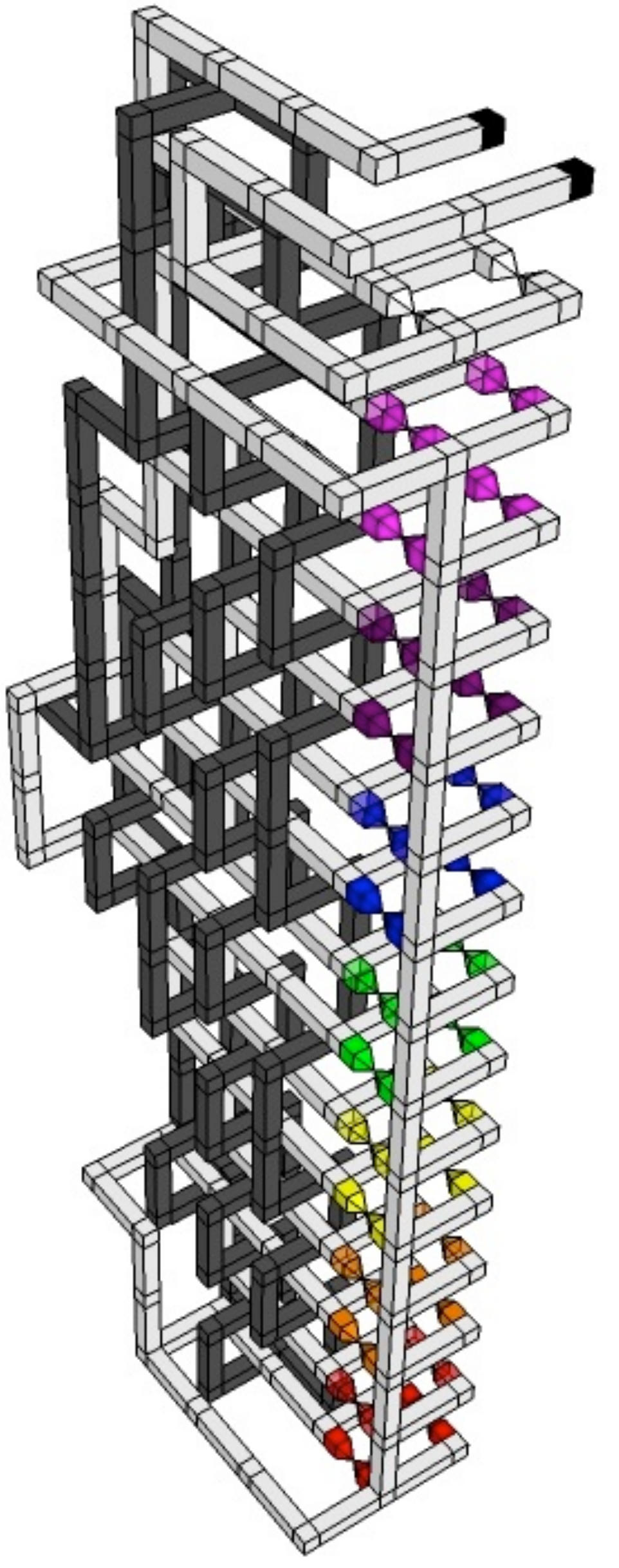}}
\end{center}
\vspace*{-15pt}
\caption{The circuit from Fig. \ref{fig:circuits}a) is rotates 90 degrees around the injection points.  This opens access to these junctions from 
the input side.  Three injection points (White, translucent pink and red) remain difficult to access due to primal defects.}
\label{fig:T-gate02}
\end{figure}

\begin{figure}[ht!]
\begin{center}
\resizebox{80mm}{!}{\includegraphics{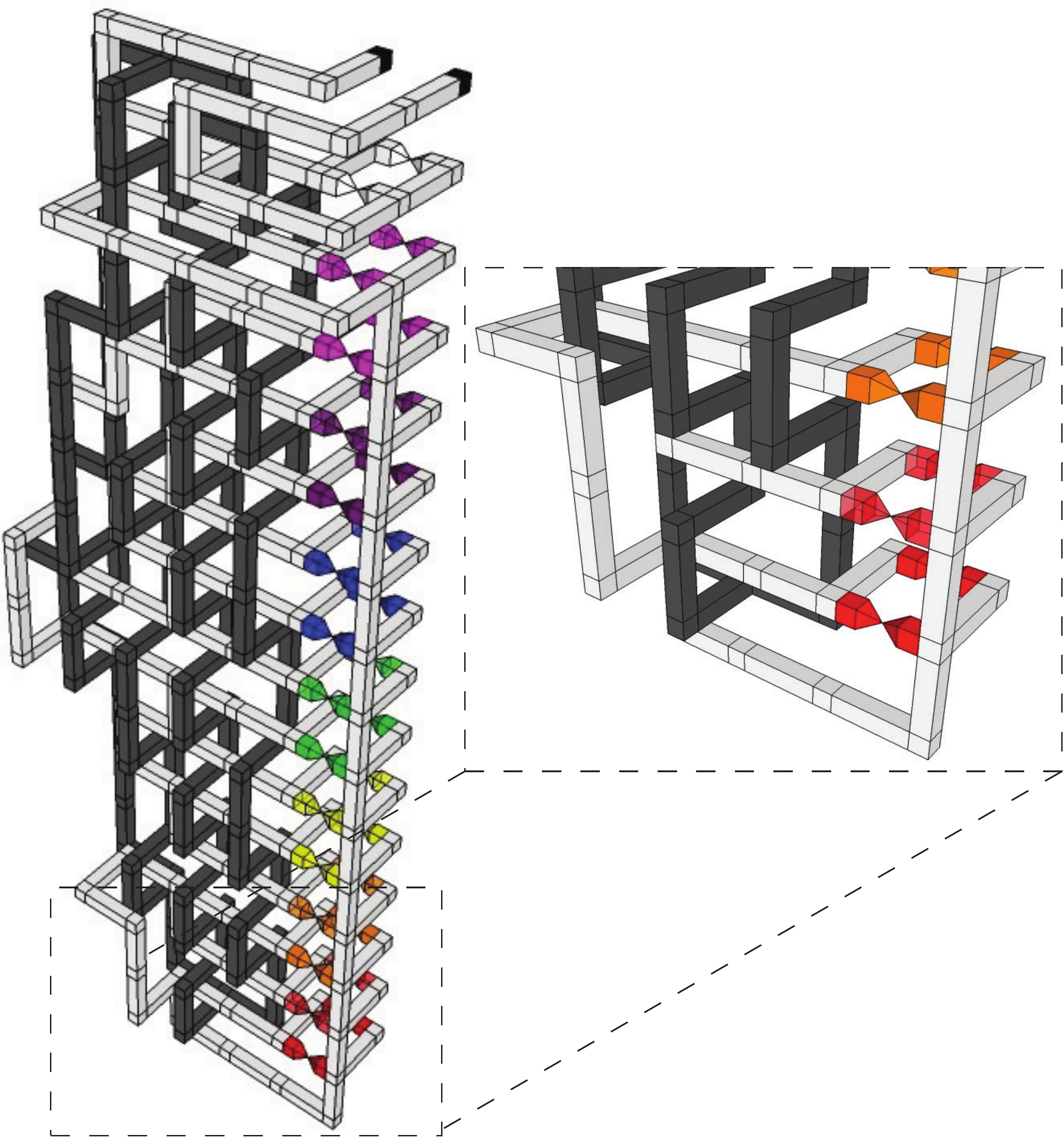}}
\end{center}
\vspace*{-15pt}
\caption{The primal defect strand near the red injection point is rotated, giving input access to the injection point}
\label{fig:T-gate03}
\end{figure}

\begin{figure}[ht!]
\begin{center}
\resizebox{80mm}{!}{\includegraphics{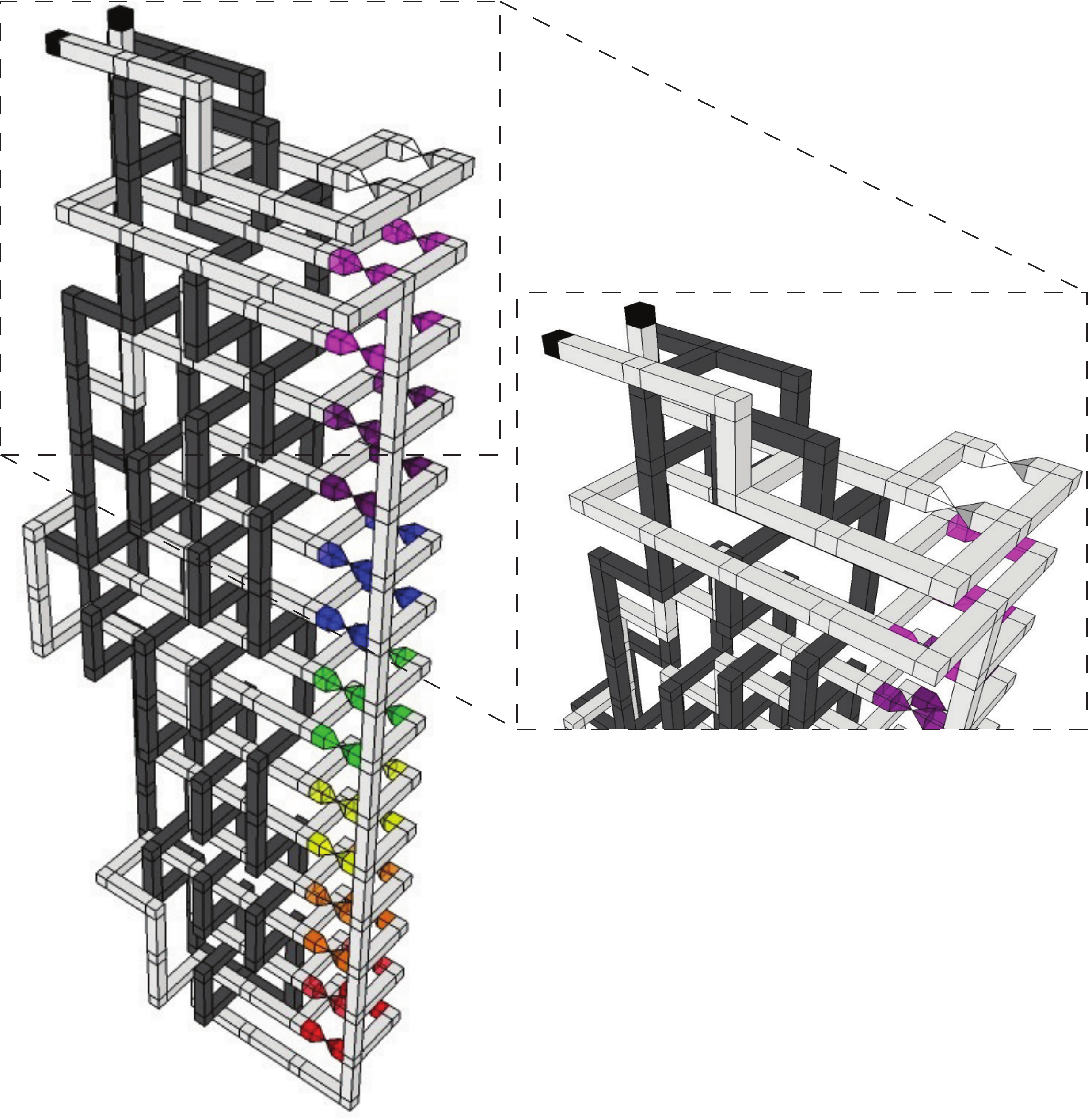}}
\end{center}
\vspace*{-15pt}
\caption{The output defects are rotated 180 degrees and moved to the left.}
\label{fig:T-gate04}
\end{figure}

\begin{figure}[ht!]
\begin{center}
\resizebox{80mm}{!}{\includegraphics{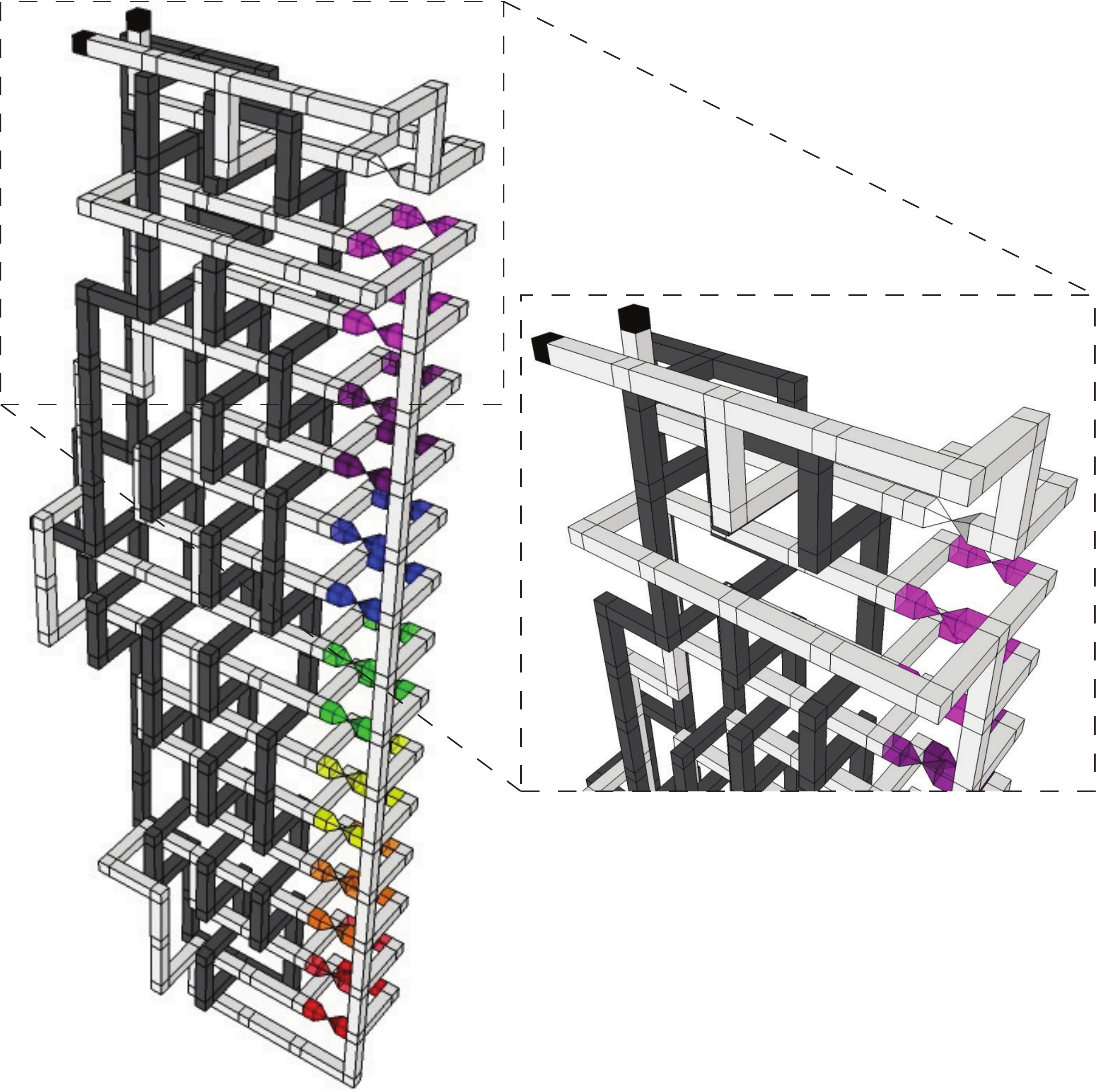}}
\end{center}
\vspace*{-15pt}
\caption{Primal defect connecting to output is raised to give access to the white injection point.}
\label{fig:T-gate05}
\end{figure}

\begin{figure}[ht!]
\begin{center}
\resizebox{45mm}{!}{\includegraphics{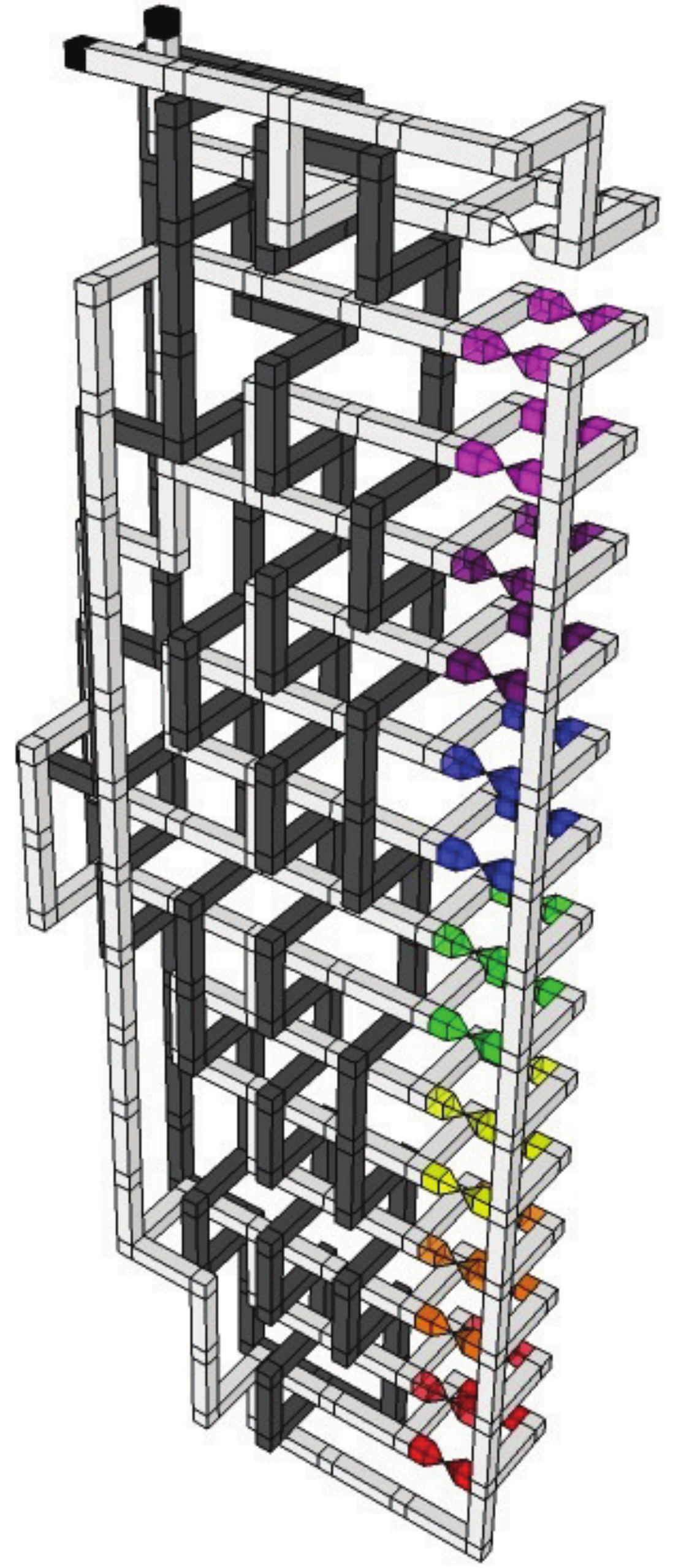}}
\end{center}
\vspace*{-15pt}
\caption{The primal defect strand blocking access to the translucent pink injection point is deformed along the rightmost primal defect 
strand creating and then further deformed to the left. This gives access to the injection point from the input side.}
\label{fig:T-gate06}
\end{figure}

\begin{figure}[ht!]
\begin{center}
\resizebox{70mm}{!}{\includegraphics{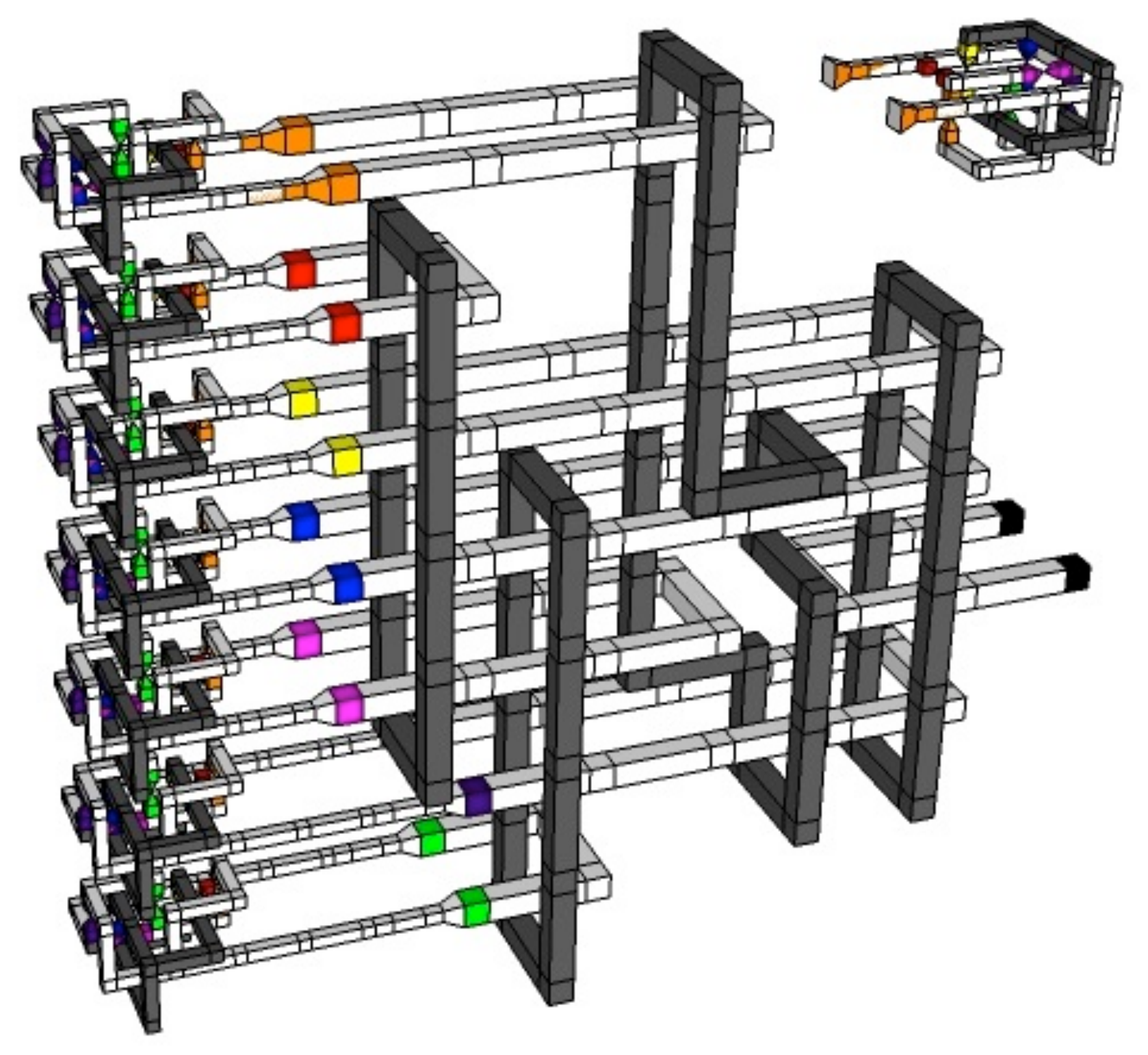}}
\end{center}
\vspace*{-15pt}
\caption{Second level distillation circuit for $\ket{Y}$ states.  In order to have clean access to the seven injection 
points at level two, a less optimised version of the circuit is used.  This structure does have the ability to be compressed further, however the majority of resources needed by the $R_z(\pi/8)$ gate is dedicated to 
$\ket{A}$ state distillation and consequently we do not compact this circuit further.  Space in the structure also exists for an additional 
level one distillation circuit to protect against circuit failure at level one.  This circuit can connect without penalty by a reordering of 
qubits in the second level circuit and connecting the auxiliary circuit to the output side.}
\label{fig:Yconcat2}
\end{figure}
These deformations allow us to access the 15 injection points that will use the output from the level one distillation circuits.  Note that the 
temporal axis of the circuit is still well defined as the injection points for the corrective $R_z(\pi/4)$ gates occur after the 
transversal $R_z(\pi/8)$ gates.  

We can now combine the structure in Fig. \ref{fig:T-gate06} with 15 copies of level one distillation circuits.  As noted in the main text and 
introduced in Ref. \cite{FD12}, 
because of the residual error associated with distillation, the error correction associated with level one does not have to be as strong as 
level two.  The size and separation of defects at level two must match up with the error correction strength at the algorithmic level, however 
the strength of error correction for the level one distillation circuits only needs to be as strong as the residual error associated with the 
circuits themselves.  Hence at level one we reduce the size and separation of defects by a factor of two.  This reduction in required error 
correction at level one allows us to stack 17 copies of Fig. \ref{fig:circuits}a) along the input edge of Fig. \ref{fig:T-gate06} and form the 
appropriate input/output connections. 
\begin{figure*}[ht!]
\begin{center}
\resizebox{180mm}{!}{\includegraphics{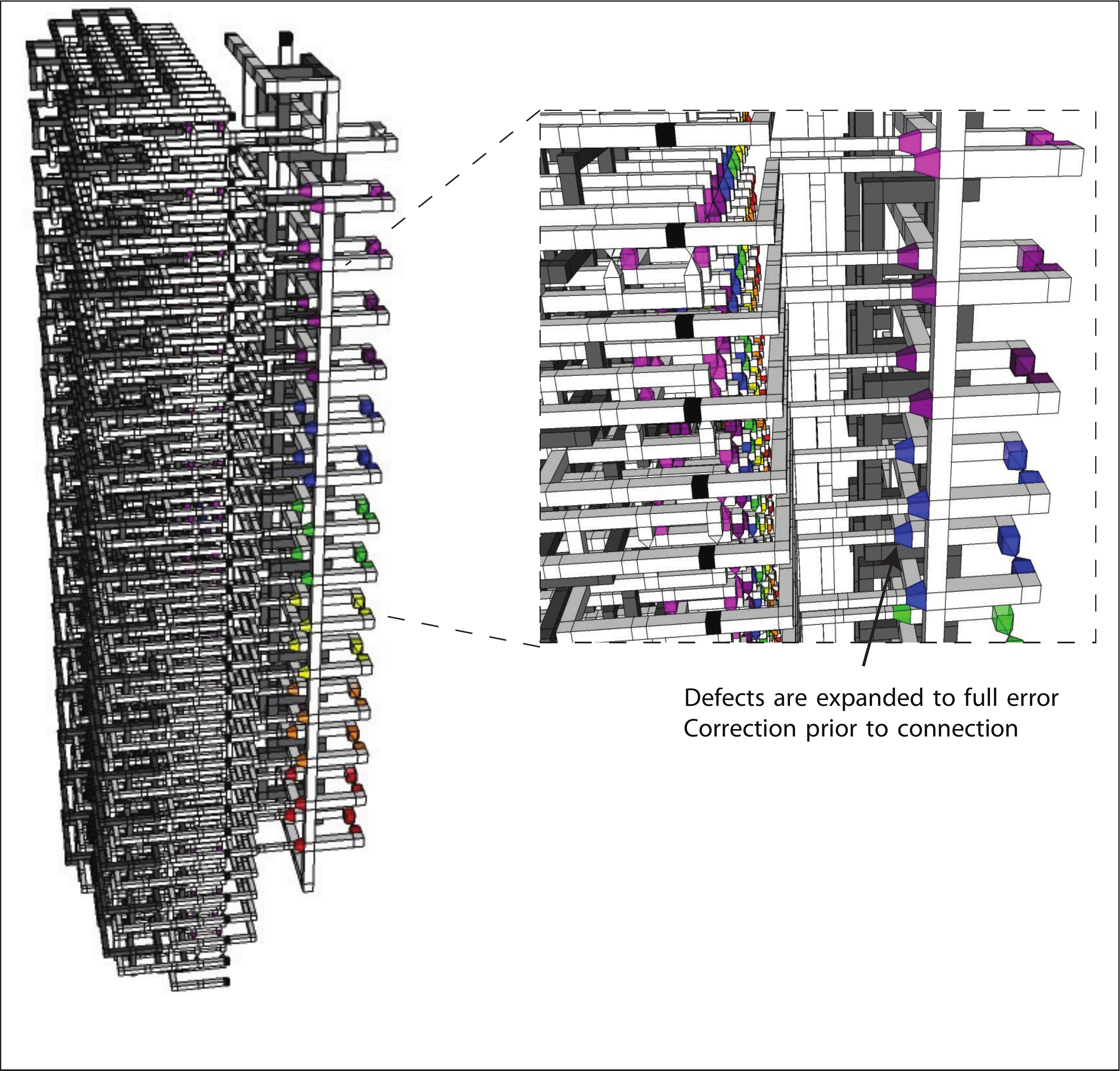}}
\end{center}
\vspace*{-15pt}
\caption{Connecting multiple first level distillation circuits to the modified second level distillation circuit.  
Before the output of level one is connected to the relevant injection points at level two the defects are expanded 
to the full strength error correction needed at the algorithmic level.  The first level distillation circuits can 
utilise a smaller error correcting code as the residual error from the distillation circuit will be higher than the 
required protection of the data qubits.  As such, the first level defects must maintain a separation from the 
second level circuits compatible with the strength of error correction at level two.}
\label{fig:concat1}
\end{figure*}
The fact that there is space to stack 17 copies of the level one distillation circuit helps us to protect agains distillation failure at level one.  
The failure of the first level to produce enough states requires three distillation circuits to fail, which occurs with a probability of order 
$\binom{17}{3} p^3$, which for $p=10^{-3}$ is $O(10^{-7})$.  Therefore, a failure at the first level of concatenation will not occur at 
every logical time step and only in certain rare times will an algorithmic qubit have to wait until a level one distillation circuit 
is redone.  

After defects are outputted from the level one circuits they are expanded to full error correction strength and attached to the level two circuit at the 
appropriate points (in Fig. \ref{fig:concat1} we have removed the pyramid structures at the injection points of the 
level two circuit, but retain the colour coding) .  
This expansion needs to be done carefully.  While the level one circuit can have a separation between defects half that 
of the level two circuit, the separation of level one defects and level two defects \emph{must} be the same as separations within the 
level two circuit.  This is because error chains can begin on a level two defect and terminate on a level one defect.

\begin{figure*}[ht!]
\begin{center}
\resizebox{180mm}{!}{\includegraphics{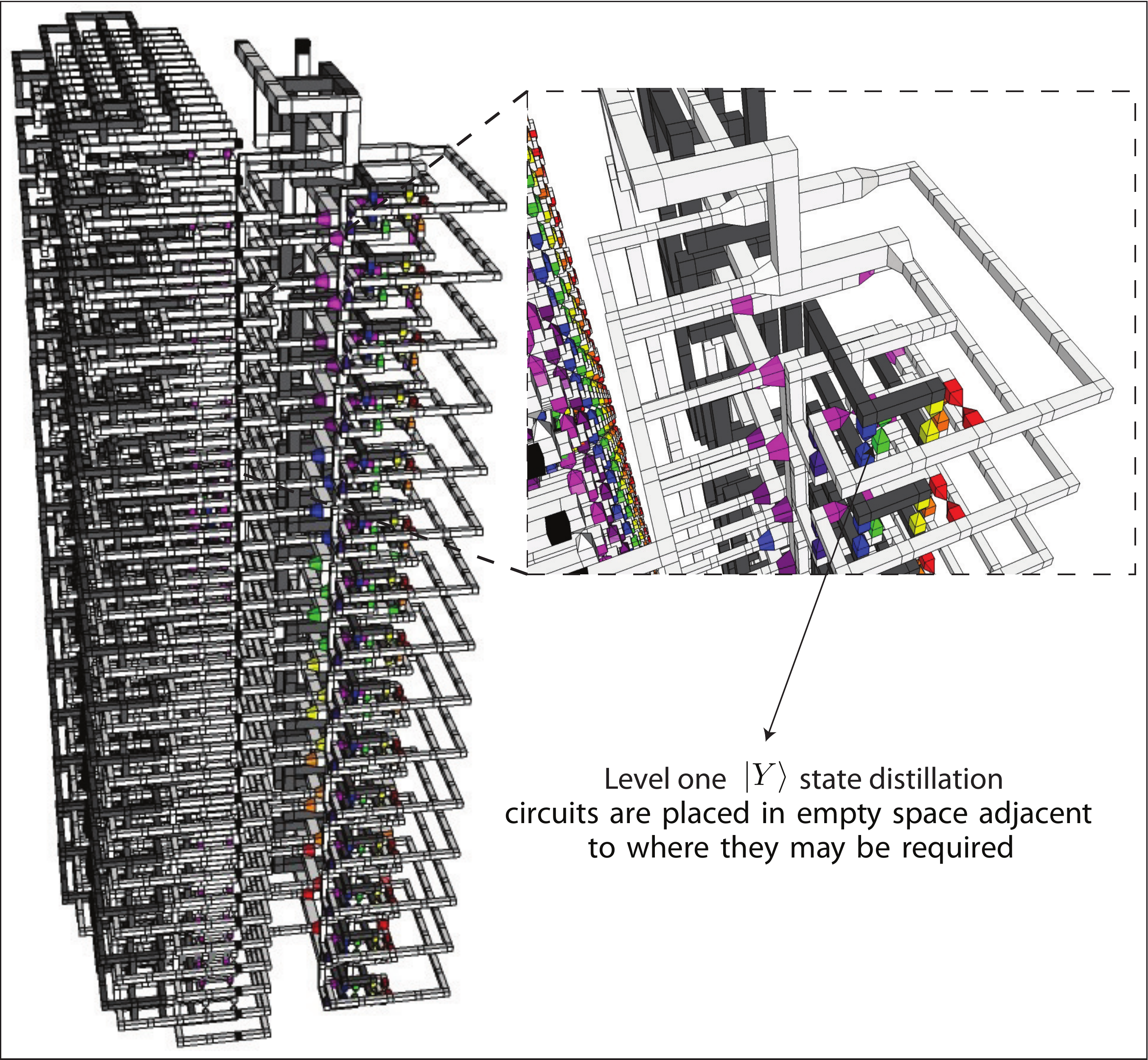}}
\end{center}
\vspace*{-15pt}
\caption{15 copies of level one $\ket{Y}$ state distillation are introduced into an empty cluster region to provide 
corrective operations to the $\ket{A}$ state distillation circuit at second level.  These circuits can utilise 
smaller defects as they are level one circuits. }
\label{fig:concat2}
\end{figure*}

The final part of this circuit is the corrective $R_z(\pi/4)$ operations that may need to be applied at the second level of concatenation.  Unlike 
level one circuits, these gates need to utilise $\ket{Y}$ states that have been distilled to level one.  Given the compact nature of the 
level one $\ket{Y}$ state circuits [Fig. \ref{fig:circuits}c)], there is sufficient space adjacent to the relevant injection points to place 15 
circuits, one for each possible $R_z(\pi/4)$ corrections of the second level 
$\ket{A}$ state circuit, Fig. \ref{fig:concat2} illustrates.
The connection points of each output of the level one $\ket{Y}$ state distillation is expanded to the appropriate separation and 
size before being joined onto the level two circuit for $\ket{A}$ state distillation.  As with level one distillation circuits for $\ket{A}$ states, 
the distillation circuits for corrective $R_z(\pi/4)$ states can fail.  With this defect arrangement, the leading order failure channel is 
when a single $\ket{Y}$ state circuit fails \textbf{and} all 15 correction gates need to be applied.  This probability is given by, 
$\frac{15p}{2^{15}}$, which for $p = O(10^{-3})$ is $O(10^{-7})$, again ensuring that additional time will only be 
needed in the computer every $O(10^7)$ logical gates.

The braiding structure of Fig. \ref{fig:concat2} now allows for the application of an encoded $R_z(\pi/8)$ gate at two levels of 
concatenated distillation, but there are still two more things to consider.  While we have introduced 15 copies of level one 
$\ket{Y}$ state distillation for the correction of the second level $\ket{A}$ state circuit, we still require a level two distilled $\ket{Y}$ state 
in order to apply a possible correction gate to the final $R_z(\pi/4)$ rotation.  As with the level one $R_z(\pi/8)$ gate, this distillation is performed 
above the algorithmic layer in the cluster.  

The second level distillation circuit for $\ket{Y}$ states is shown in Fig. \ref{fig:Yconcat2}.  Note that we have used two different circuits 
for level one distillation [Fig. \ref{fig:circuits}c)] and level two [Fig. \ref{fig:circuits}b)].  We have not attempted to perform further 
compression of the level two circuit manually as its volume is sufficiently small as to not impact the depth of the overall $R_z(\pi/8)$ 
gate.  

This circuit can now be incorporated into the larger structure, with the appropriate SWAP space left between the algorithmic layer and 
the $\ket{Y}$ state distillation layer. 
\begin{figure*}[ht!]
\begin{center}
\resizebox{100mm}{!}{\includegraphics{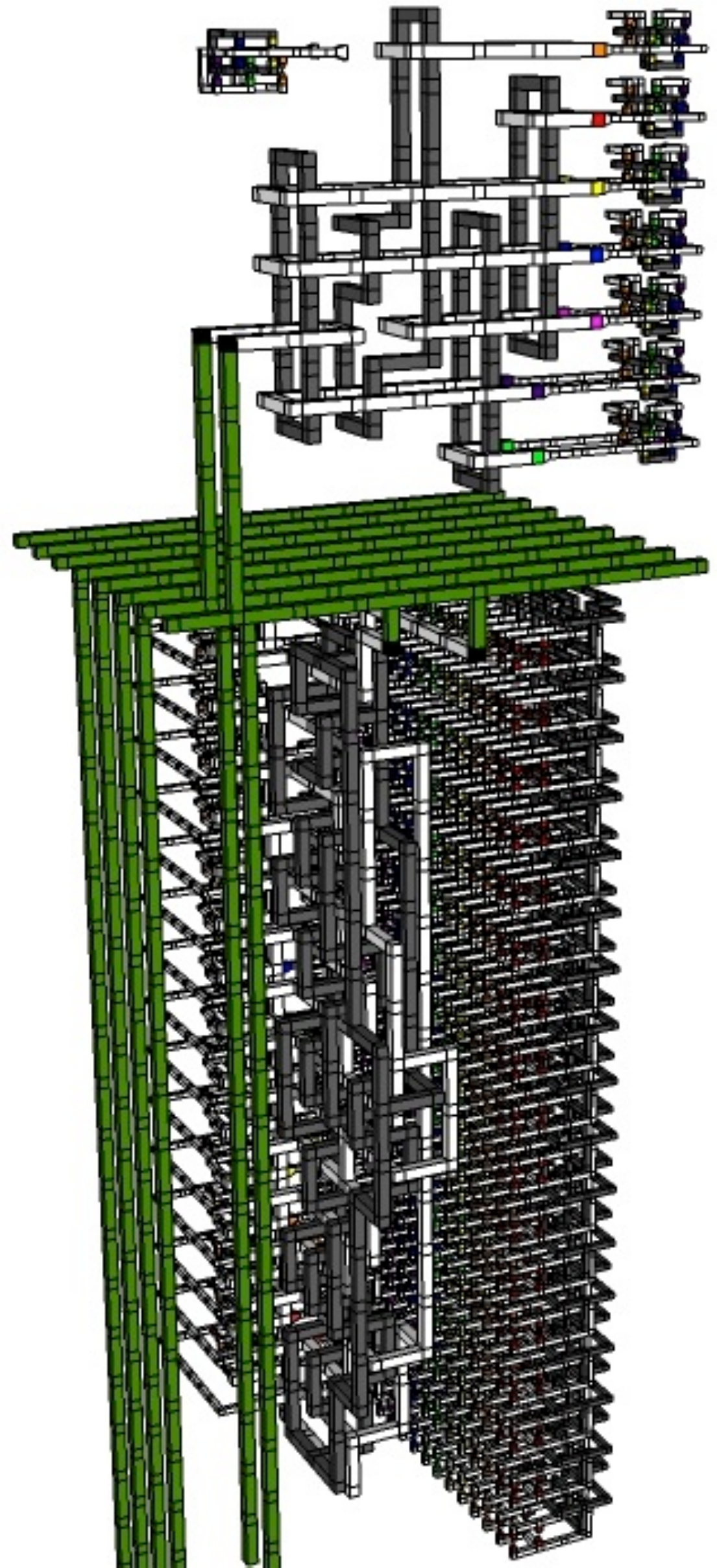}}
\end{center}
\vspace*{-15pt}
\caption{Complete structure for the $R_z(\pi/8)$ rotation for one of the four algorithmic qubits present in the repeating unit (we have 
reversed the temporal axis in this image for readability).  Two 
levels of $\ket{A}$ state distillation, with necessary correction gates sit below the algorithmic layer and two levels of $\ket{Y}$ state 
distillation exists above the algorithmic layer for the final $R_z(\pi/4)$ correction, required 50\% of the time.  
There are additional redundant circuit elements that are included to protect against the failure of level one distillation circuits.  
The connection structures illustrated here assume no such failures occur.  This circuit would have to be modified dynamically 
depending on the result of certain measurements. }
\label{fig:Yconcat3}
\end{figure*}
Along with the second level $\ket{Y}$ state distillation circuit, we have illustrated where an additional first level circuit can be placed 
in order to compensate for the possibility that one of the seven, first level circuits fail.  These failure are also compensated by the 
fact that the final second level $\ket{Y}$ states are only needed 50\% of the time.  Hence this circuit element, utilised every time a 
$R_z(\pi/8)$ gate is applied, over supplies distilled $\ket{Y}$ states.

From Fig. \ref{fig:Yconcat3}, the last issue to solve should be clear.  The space utilised for $\ket{A}$ state distillation occupies a cross 
sectional space in the lattice equal to four algorithmic qubits.  Therefore, we need to duplicate the distillation structure vertically in 
order to produce sufficient states to serve these four algorithmic qubits.  Stacking three additional copies of the $\ket{A}$ state 
distillation circuits below the one in Fig. \ref{fig:Yconcat3} gives us the stackable braiding sequence which enacts $R_z(\pi/8)$ gates 
over four algorithmic qubits.  This leads to the final structure illustrated in the main text.  

Implicit in these images (and throughout the discussion) is the possibility of dynamical configuration of these structures if level one 
distillation circuits fail.  These diagrams assume that all distillation circuits output successfully and can be connected as shown.  If this 
is not the case, a reconfiguration of the overall circuit is needed.  The design of these reconfigured circuits can be done 
offline, but their actual application will be chosen dynamically as the computation is run.  
As discussed, we have given sufficient space within the second level structure to ensure 
that a total failure (i.e. one where we do not have sufficient distilled states at the first level of concatenation) does not occur 
at every time step of computation.  However, it is expected that at least \emph{one} circuit will fail throughout the computer, at every 
logical time step, that \emph{can} be compensated by these extra resources.  The dynamical reconfiguration is not expected to increase 
the depth of these fault-tolerant gates in a significant way, but do still need to be calculated.  Given the large number of possible failure points 
the specification of all possible configurations will need to be done in an automated manner and is the focus of future work.

\end{document}